\documentclass[prd,twocolumn,floatfix,preprintnumbers,letterpaper]{revtex4}
\usepackage{graphicx}
\usepackage{amsmath,amssymb}
\usepackage{epsf}

\begin{document}

\def\be{\begin{equation}}
\def\ee{\end{equation}}

\title{Phantom Dark Energy Models with a Nearly Flat Potential}
\author{Robert J. Scherrer}
\affiliation{Department of Physics and Astronomy, Vanderbilt University,
Nashville, TN  ~~37235}
\author{A.A. Sen}
\affiliation{Center For Theoretical Physics,
Jamia Millia Islamia, New Delhi 110025, India}

\begin{abstract}
We examine phantom dark energy models produced
by a field with a negative kinetic term and
a potential that satisfies the slow roll conditions:  $[(1/V)(dV/d\phi)]^2 \ll 1$
and $(1/V)(d^2 V/d\phi^2) \ll 1$.
Such models provide a natural mechanism to produce
an equation of state parameter, $w$, slightly less than $-1$ at present.
Using techniques previously applied to quintessence,
we show that in this limit, all such phantom models converge
to a single expression for $w(a)$, which is a function
only of the present-day values of $\Omega_\phi$ and $w$.
This expression is identical to the corresponding behavior
of $w(a)$ for quintessence models in the same limit.
At redshifts $z \alt 1$, this limiting behavior
is well fit by the linear parametrization, $w=w_0 + w_a(1-a)$,
with $w_a \approx -1.5(1+w_0)$.
\end{abstract}

\maketitle

\section{Introduction}

Observations \cite{Knop,Riess} indicate
that approximately 70\% of the energy density in the
universe is in the form of an exotic, negative-pressure component,
dubbed dark energy.  (See Ref. \cite{Copeland}
for a recent review). Taking
$w$ to be the ratio of
pressure to density for the dark energy:
\begin{equation}
w = p_{DE}/\rho_{DE},
\end{equation}
recent observations suggest that $w$ is close to $-1$.
For example, if $w$ is assumed to be constant, then
$-1.1 \alt w \alt -0.9$ \cite{Wood-Vasey,Davis}.  If we are interested in dynamical
models for dark energy, such as those that arise from a scalar
field $\phi$, then such models can be significantly simplified in
the limit that $w$ is close to $-1$.  This fact was exploited
in Ref. \cite{SS}, which examined quintessence models
with a nearly flat potential, defined as a $V(\phi)$ satisfying
the slow-roll conditions:
\begin{equation}
\label{slow1}
\left(\frac{1}{V} \frac{dV}{d\phi}\right)^2 \ll 1,
\end{equation}
and
\begin{equation}
\label{slow2}
\frac{1}{V}\frac{d^2 V}{d\phi^2} \ll 1.
\end{equation}
With these assumptions, it is possible to derive a generic
expression for $w$ as a function of the scale factor, $a$,
that provides an excellent approximation to this entire class of potentials.  (For other
approaches to this problem, see Refs. \cite{Crit,Neupane,Cahn}).

Here we extend these results to the case of phantom models,
i.e., models for which $w < -1$.
It was first noted by Caldwell \cite{Caldwell} that
observational data do not rule out the possibility
that $w <-1$.  These phantom dark energy
models have several interesting properties.
The density of the dark energy increases with
increasing scale factor,
and the phantom energy density can become infinite
at a finite time, a condition known as the ``big rip" \cite{Caldwell,rip,rip2}.
Further, it has been suggested that the finite lifetime for the universe
which is exhibited in these models may provide an explanation for
the apparent coincidence between the current values of the
matter density and the dark energy density \cite{doomsday}.  (See
Ref. \cite{Lineweaver} for a related, but different argument).

A simple way to achieve a phantom model is to take a scalar
field Lagrangian with a negative kinetic term.  Such
models have well-known problems \cite{Carroll,Cline,Hsu1,Hsu2},
but they
nonetheless provide an interesting set of representative phantom
models, and they have been widely studied \cite{Guo,ENO,NO,Hao,Aref,Peri,Sami,
Faraoni,Chiba,KSS}.  Here we assume a negative kinetic
term, and then use
techniques similar to those in Ref. \cite{SS} to derive an expression for $w(a)$
for phantom models satisfying conditions (\ref{slow1}) and (\ref{slow2}).

We assume a spatially-flat universe containing only
nonrelativistic matter and phantom dark energy, since
radiation can be neglected in the epoch of interest.
Then
\be
\label{Fried}
H^2 = {\rho_T}/{3},
\ee
where $H = \dot a/a$, $a$ is the scale factor,
$\rho_T$ is the total energy density, 
and we take $\hbar = c = 8\pi G = 1$ throughout.

In a phantom model with negative kinetic
term and potential $V(\phi)$, the energy density
and pressure of the phantom are given by
\be
\rho_{\phi}={-(1/2)\dot \phi^2 + V(\phi)},
\ee
and
\be
p_{\phi}={-(1/2)\dot \phi^2 - V(\phi)},
\ee
so that the equation of state parameter is
\begin{equation}
\label{w}
w =\frac{(1/2)\dot \phi^2 + V(\phi)}{ (1/2)\dot \phi^2 - V(\phi)}.
\end{equation}

The evolution of $\phi$ is given by
\begin{equation}
\label{phiev}
\ddot{\phi}  +   3 H \dot{\phi} - V^\prime(\phi) = 0.
\end{equation}
where the prime denotes the derivative with respect
to $\phi$.
A field evolving according to equation (\ref{phiev}) rolls
uphill in the potential.

This equation of motion can be rewritten as \cite{Chiba,KSS}
\begin{equation}
\label{Vprime/V}
\pm \frac{V^\prime}{V} = \sqrt{\frac{-3(1+w)}{\Omega_\phi}}\left[1 + \frac{1}{6}
\frac{d\ln(x)}{d\ln(a)}\right],
\end{equation}
where
$\Omega_\phi$ is the density of the
phantom field in units of the critical
density (note that $\Omega_\phi$ evolves with time).
In equation (\ref{Vprime/V}), $x = \dot\phi^2/2V$, so that
$x$
and $w$ are related via
\begin{equation}
x = - \frac{1+w}{1-w}.
\end{equation}
(Note that in equation (\ref{Vprime/V}) we use
the sign convention in \cite{KSS}, rather than the one used in \cite{Chiba}).
Equation (\ref{Vprime/V}) is the phantom
version of the quintessence equation of motion first derived
in Ref. \cite{Steinhardt}; it differs from the quintessence equation
only in the sign of $1+w$ on the right-hand side.

We are interested in the limit where $w$ is near $-1$, so following
Ref. \cite{SS},
we define
$\beta \equiv -(1+w)$,
where we take $\beta$ to be small and positive.  This will allow us to
drop terms of higher order in $\beta$.  Then equation (\ref{Vprime/V})
becomes
\begin{equation}
\label{betaprime}
\beta^\prime = -3\beta(2+\beta) + \lambda (2+\beta)\sqrt{3 \beta \Omega_\phi},
\end{equation}
where we have defined $\lambda = V^\prime/V$.
The fractional density in phantom dark energy, $\Omega_\phi$, evolves
as
\begin{equation}
\label{Omegaprime}
\Omega_\phi^\prime = 3(1+\beta)\Omega_\phi (1-\Omega_\phi). 
\end{equation}
Combining equations (\ref{betaprime}) and (\ref{Omegaprime})
yields
\begin{equation}
\label{exact1}
\frac{d\beta}{d\Omega_\phi} = \frac{-3\beta(2+\beta) + \lambda (2+\beta)\sqrt{3 \beta
\Omega_\phi}}{3(1+\beta)\Omega_\phi (1-\Omega_\phi)}
\end{equation}
This can be compared to the corresponding equation for $d\gamma/d\Omega$
for quintessence, where $\gamma = 1+w$ \cite{SS}:
\begin{equation}
\label{exact2}
\frac{d\gamma}{d\Omega_\phi}
= \frac{-3\gamma(2-\gamma) + \lambda(2-\gamma)\sqrt{3 \gamma \Omega_\phi}}
{3(1 - \gamma)\Omega_\phi(1-\Omega_\phi)}.
\end{equation}

Clearly, equations (\ref{exact1}) and (\ref{exact2}) predict that $\beta(\Omega_\phi)$
for a phantom field and $\gamma(\Omega_\phi)$ for a quintessence field will evolve quite
differently.  Now, however, we make our slow-roll assumptions.  As in Ref. \cite{SS},
we take $\beta \ll 1$ and we assume $\lambda$ to be a constant, $\lambda_0$, given by
the initial value of $V^\prime/V$.  Both of these results are a consequence of
equations (\ref{slow1}) and (\ref{slow2}); see Ref. \cite{SS} for the details.
Then equation (\ref{exact1}) becomes
\begin{equation}
\label{step2}
\frac{d\beta}{d\Omega_\phi} = -\frac{2\beta}{\Omega_\phi(1-\Omega_\phi)} +
\frac{2}{3}\lambda_0\frac{\sqrt{3\beta}}{(1-\Omega_\phi)\sqrt{\Omega_\phi}}.
\end{equation}
This is identical to the equation one obtains, in the corresponding limit,
for $\gamma(\Omega_\phi)$ for quintessence \cite{SS}.  Thus,
in this limit, $1+w(\Omega_\phi)$ for quintessence and $-1-w(\Omega_\phi)$
for a phantom field evolve in exactly the same way.  Using the exact
solution for equation (\ref{step2}) from Ref. \cite{SS}, we obtain
\begin{equation}
\label{wOm}
1+w = -\frac{\lambda_0^2}{3}\left[\frac{1}{\sqrt{\Omega_\phi}}
- \frac{1}{2}\left(\frac{1}{\Omega_\phi} - 1 \right)
\ln \left(\frac{1+\sqrt{\Omega_\phi}}
{1-\sqrt{\Omega_\phi}} \right)\right]^2.
\end{equation}
Further, when $1+w$ is close to $-1$, we can approximate the solution 
to equation (\ref{Omegaprime}) as \cite{SS,Crit}
\begin{equation}
\label{Oma}
\Omega_\phi = \left[1 + \left(\Omega_{\phi 0}^{-1} - 1 \right)a^{-3}
\right]^{-1},
\end{equation}
where $\Omega_{\phi 0}$ is the present-day
value of $\Omega_\phi$, and we take $a=1$ at the present.
Combining equations (\ref{wOm}) and (\ref{Oma}), and normalizing to $w = w_0$
at present, we obtain:
\begin{eqnarray}
1 + w = (1+ w_0)\Biggl[ \sqrt{1 + (\Omega_{\phi 0}^{-1} - 1)a^{-3}}\nonumber\\
- (\Omega_{\phi 0}^{-1} - 1)a^{-3} \tanh^{-1} \frac{1}{\sqrt{1 + (\Omega_{\phi 0}^{-1}-1)a^{-3}}}
\Biggr]^2\nonumber\\
\label{wpred}
\times \left[\frac{1}{\sqrt{\Omega_{\phi 0}}}
- \left(\frac{1}{\Omega_{\phi 0}} - 1 \right)
\tanh^{-1}\sqrt{\Omega_{\phi 0}}\right]^{-2}.
\end{eqnarray}
This equation is identical to the corresponding result for $1+w$ for
the case of quintessence \cite{SS}.  The fact
that these two different models (phantom and quintessence)
yield an identical form for $(1+w)/(1+w_0)$ is not {\it a priori} obvious;
indeed, this result is valid only in the ``slow roll" limit
considered here.

\begin{figure}[t]
\centerline{\epsfxsize=3.7truein\epsfbox{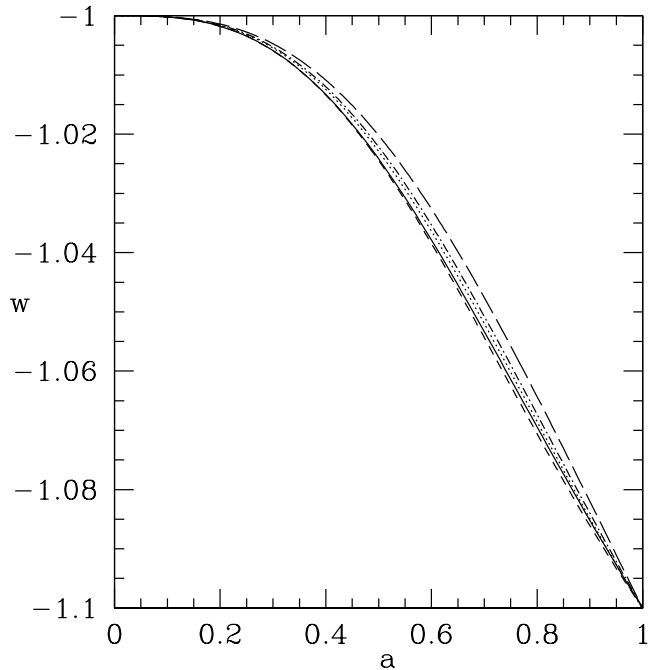}}
\caption{The evolution of $w$ for the phantom field evolving
in a nearly flat potential as a function of the scale factor,
$a$, normalized to $a=1$ at the present, with $\Omega_{\phi 0} = 0.7$
and $w_0 = -1.1$.  Solid
curve is our analytic result for the behavior
of phantom models with a nearly flat potential and $w$ near $-1$.  Other curves
give the true evolution for the potentials
$V(\phi) = \phi^6$ (dotted), $V(\phi) = \phi^{2}$ (short dash),
$V(\phi) = \phi^{-2}$ (long dash),
and $V(\phi) = \exp(\lambda\phi)$ (dot-dash).}
\end{figure}
\begin{figure}[t]
\centerline{\epsfxsize=3.4truein\epsfbox{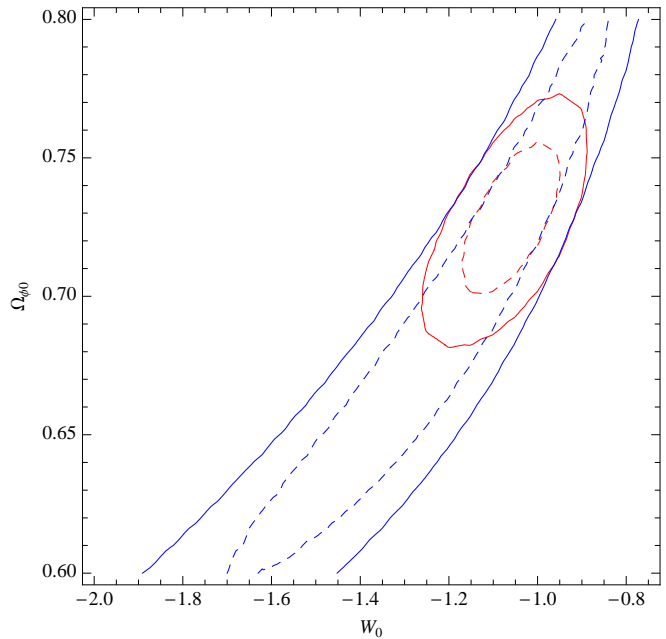}}
\caption{The $1-\sigma$ (solid) and $2-\sigma$ (dashed)
contours in the plane defined
by the present-day values of $\Omega_\phi$ and $w$ (denoted $\Omega_{\phi 0}$
and $w_0$ respectively) for
the phantom ($w_0 < -1$) or quintessence ($w_0 > -1$) field, in the limit
of a nearly flat potential.  Blue (outer) contours are limits
from SNIa alone.  Red (inner) contours are SNIa+BAO.}
\end{figure}

We now compare the approximation given in equation (\ref{wpred}) to
exact numerical results.  Kujat et al. \cite{KSS} showed that phantom
potentials with a negative kinetic term can be divided into three broad
classes, depending on the late-time asymptotic behavior of $V^\prime(\phi)/V(\phi)$.
When
$V^\prime/V \rightarrow \pm \infty$ at late times, $w_\phi \rightarrow -\infty$.
These models include, for instance, negative power laws.  For
models in which
$V^\prime/V$ asymptotically approaches a constant, $w$ also
approaches a constant, $w_0$, where $w_0 < -1$.  This
class of models includes exponential potentials.
Finally, when $V^\prime/V \rightarrow 0$ at late times, $w_\phi \rightarrow -1$.  This
class of models includes, for example, positive power law potentials.
This final class of models, for which $w \rightarrow -1$,
can be further subdivided into models which yield a future singularity (a ``big rip"), and those
which do not.  The exact conditions necessary to avoid a big rip
involve an integral function of $V(\phi)$ \cite{KSS}, but
for power-law potentials, the condition is simpler:  for $V(\phi) =
\phi^\alpha$, $\alpha > 4$ yields a big rip, while
$\alpha \le 4$ yields no future singularity \cite{Sami,KSS}.

We have therefore chosen to compare our analytic approximation for $w(a)$
with a model from each of these four classes.  The results are
displayed in Fig. 1.
As in the case of quintessence, the agreement between
the actual evolution and our approximation in equation (\ref{wpred})
is remarkably good, with the error in $w$ less than 0.5\%.
Our analytic approximation is well-fit by a linear
relation, \cite{CPL}
\begin{equation}
w(a) = w_0 + w_a(1-a),
\end{equation}
in the regime $a > 0.5$, which corresponds to a redshift $z < 1$.
Further, for the value of $\Omega_{\phi0}$ adopted here ($\Omega_{\phi0}=0.7$),
we have $w_a \approx -1.5(1+w_0)$.  This is the same
relation between $w_a$ and $w_0$ noted empirically for linear
quintessence potentials \cite{Kallosh}, predicted for
generic slow-roll quintessence models \cite{SS}, and derived
analytically for linear quintessence potentials \cite{Cahn}.
(Note that the latter reference gives a value of $-1.3$ rather
than $-1.5$, but this difference is not significant, as the result
depends on the assumed value of $\Omega_{\phi 0}$ and on the
assumed definition of $w_a$; the latter is ambiguous
since $w(a)$ is not exactly linear in $a$ in these models).

Given that equation (\ref{wpred}) provides a generic prediction for the behavior of
$w(a)$ in a wide class of phantom models, it is useful to compare this prediction
to the observations.  Further, since both quintessence and phantom models
produce the same generic form for $w(a)$ in the limit where $V^\prime/V$ is small
and nearly constant, we can combine the results for $w(a)$ for quintessence
models from Ref. \cite{SS} with the results for $w(a)$ for phantom models
given here into a single likelihood plot.
In Fig. 2, we compare this generic prediction for $w(a)$
to the SNIa and baryon acoustic oscillation data.
The likelihoods were constructed using the 60 Essence supernovae, 57 SNLS (Supernova
Legacy Survey) and 45 nearby supernovae, and the
data release of 30 SNe Ia detected by HST and classified as the Gold sample by
Riess et al. \cite{Riess}.  The combined dataset can be found in Ref. \cite{Davis}.
We also add the measurement of the baryon acoustic oscillation (BAO)
scale at $z_{BAO} = 0.35$ as observed by the
Sloan Digital Sky Survey \cite{sdss}.

It is important to note that Fig. 2 combines two completely different models.  The
portion of the figure with $w_0 > -1$ corresponds to quintessence models, while
$w_0 < -1$ corresponds to phantom models.  We have simply taken advantage
of the fact that both models yield the same form for $w(a)$ as a function
of $w_0$ and $\Omega_{\phi0}$ in the limiting case where the potential is nearly flat.

Our main conclusion is that the analytic approximation previous derived
for quintessence with a nearly flat potential \cite{SS} has the
same form when it is extended to phantom models with a negative
kinetic term.  Thus, the generic prediction for the evolution $w$
at low redshift ($z \alt 1$)
in such
models, $w(a) = w_0 -1.5(1+w_0)(1-a)$, can be applied to both
$w_0 > -1$ and $w_0 < -1$.

\acknowledgements 

R.J.S. was supported in part by the Department of Energy (DE-FG05-85ER40226). A.A.S acknowledges the financial support from the University Grants Commission, Govt. of India through Major Research project Grant (Project No: 33-28/2007(SR)).

\end{document}